\documentstyle[12pt]{article}
         \makeatletter
         \def\thefigure{\@arabic\c@figure}\def\fps@figure{tbp}
         \def\ftype@figure{1}\def\ext@figure{lof}
         \def\fnum@figure{\protect\footnotesize Fig.\ \thefigure}
         \def\thetable{\@arabic\c@table}
         \def\fps@table{tbp}\def\ftype@table{2}\def\ext@table{lot}
         \def\fnum@table{\protect\footnotesize Table \thetable}
         \makeatother
\begin{document}
\vspace*{0.3in}
\begin{center}
  {\Large \bf
Probing the ground state and transition densities of halo
nuclei}\\
  \bigskip
  \bigskip
  {\Large  C.A. Bertulani$^{(a,c)}$ and H. Sagawa$^{(b)}$}  \\
  \bigskip

$(a)$  Gesellschaft f\"ur Schwerionenforschung, KPII\\
   Planckstr. 1, D-64291 Darmstadt, Germany\\

 \bigskip

$(b)$  Center for Mathematical Sciences, University of Aizu\\
   Ikki-Machi, Aizu-Wakamatsu, Fukushima, 965,   Japan  \\

  \end{center}

\bigskip
\centerline{\bf ABSTRACT}
\begin{quotation}
\vspace{-0.10in}
We investigate the use of elastic and inelastic
scatterings with secondary beams of radioactive
nuclei as a mean to obtain information on ground
state properties and
transition matrix elements to continuum states.
An eikonal model is developed for this purpose
by using the folding potential.
In particular we discuss possible
signatures of halo wavefunctions in elastic and inelastic
scattering experiments.

\end{quotation}

\vfil

(c) Present address: Instituto de F\'\i sica, UFRJ, 21945-970 Rio de Janeiro,
Brazil

\newpage
\baselineskip 4ex

\section{Introduction}

Reactions with radioactive beams are a useful tool to understand the
properties of nuclei far from the stability line (see, e.g. ref.
\cite{Ka91}).
Up to the present
these reactions are mainly restricted to studies of reaction
cross sections and momentum distributions of the fragments
\cite{Tan85,Tan88}.
Elastic scattering has been studied in several
experiments \cite{Mo92,Ko93,Ko92,Za94,Tak92}. Inelastic
scattering has also been studied, but very few experiments for
Coulomb breakup processes \cite{Iek93,Nak94} only are available so far.
Usual techniques with stable nuclei, like photo-nuclear absorption and
electron scattering are far beyond the present experimental
possibilities.
The study of reaction cross sections, momentum distributions,
and elastic scattering gives us a very limited access to the
information on the internal structure of exotic nuclei, although
many intriguing properties of these nuclei have been deduced in
these experiments.
An example is the discovery of the extended matter distribution
in very light neutron-rich nuclei, so-called halo nuclei \cite{Tan85}.

The study of inelastic excitation cross sections is the natural step
to increase our knowledge on the nuclei far from the stability
line.
In fact, Coulomb and nuclear excitations in nucleus-nucleus scattering
are well established tools for the spectroscopy of stable nuclei and
are complementary to photo-nuclear and electron scattering experiments.
Due to the low luminosity of radioactive nuclear beams,
experiments are possible only when the
cross sections are sufficiently large. For inelastic scattering this
is the case
for the Coulomb excitation of loosely-bound nuclei, e.g., $^{11}Li$
and $^{11}Be$, incident on heavy targets \cite{Iek93,Nak94}.
The beams of exotic nuclei are often available
at intermediate and high energies, $E_{lab}>50$ MeV/nucleon \cite{Ka91}.
At these energies the Coulomb field favors the dipole excitations.
On the other hand, the nuclear interaction favors the monopole and
quadrupole excitations.

A study of nuclear excitations of halo  nucleus $^{11}Li$ has been
done in ref. \cite{BS93}. It was shown that the excitation of low-lying
continuum states of $^{11}Li$ have substantial
cross sections and
it  can reach some 100  mb/sr at forward angles.
In this article we extend the model of ref. \cite{BS93} to include
Coulomb excitation processes. Inelastic excitation processes are
unavoidably accompanied by elastic scattering. We thus also make a
study of elastic scattering processes with unstable nuclei.
Under simplifying assumptions, which we try to justify in the next
sections, elastic scattering
allows one to get information on the ground state properties
of the exotic nuclei, while the inelastic scattering processes tells
us about the transition densities for the excitation.
If the ground state wave function can be
deduced from the elastic scattering,
the excited state wavefunctions
are possible to infer from the inelastic excitation processes.

The extraction of the above mentioned information
from the experimental data suffers from several
difficulties which arise due to the complexity of the reaction mechanism.
The ideal situation occurs when the reaction mechanism can be explained
with very simple models. In this article we use
a simple and tractable model for the
reaction mechanisms and apply it to the study of the scattering
of unstable beams. Under some circumstances it is shown that our
model can be very useful to extract the ground state properties of
exotic nuclei. A study of this feature is presented in sections 3 and 4.
A brief explanation of the
calculation ground state densities for exotic nuclei is done in
section 3.  These densities are
used as inputs in the calculation of the elastic cross sections, which are
compared to some available experimental data
in section 4. To show the feasibility
of the models used, we also compare the theoretical predictions
with some experimental data for
reactions with stable nuclei.

The inelastic scattering cross sections are discussed in sections
5 and 6. In section 5 we deduce formulas using the deformed
potential model and folding model for inelastic excitations.
The equations for the folding  model has been used in ref. \cite{BS93}.
We also present  appropriate formulas for the case of
Coulomb excitation.  These equations are applied in section 6 to
discuss the angular distributions for inelastic processes with
stable and unstable nuclear beams.
In section 7 we present our conclusions.

\bigskip\bigskip

\section{Elastic Scattering}

The elastic scattering in nucleus-nucleus collisions
is a well established tool for the investigation of
ground state densities. This is because
the optical  potential can be related to the ground
state densities by means of a folding of the
nucleon-nucleon  interaction with the  nuclear
densities of two colliding nuclei. But,
this relationship is not quite straightforward. It depends
on the effective interaction used, a proper treatment of
polarization effects, and so on  (for a review see, e.g., \cite{HRB91}).
At higher bombarding energies than $E_{Lab} \sim 50$ MeV/nucleon,
a direct relationship
between the nuclear densities and the optical potential is
possible, as long as the effects of multiple nucleon-nucleon scattering
can be neglected. The effects of real, or virtual,
nuclear excitations are small since the
excitation energies involved  are much smaller than the
bombarding energies. Lenzi, Vitturi and Zardi \cite{LVZ88}
have performed an extensive study of the nuclear scattering
of stable nuclei at high energies. From their study one concludes
that a simple relationship between the ground state densities
and the elastic scattering cross sections quite often
yields
very reasonable results, as compared to the experiments.
We adopt here a similar approach and extend it
to study the scattering of exotic nuclei.

At high energies the elastic
scattering cross section for proton-nucleus collisions is
well described by means of the eikonal approximation
\cite{Gl59}. The optical
potential for proton-nucleus scattering is assumed to be of the form
\begin{equation}
U(r)=U_0(r)+U_S(r)\ ({\bf L.S})+U_C(r)
\end{equation}
where
\begin{equation}
U_0(r)=V_R\ f_R(r)-iW_V\ f_I(r)+4\ i \ a_I \ W_I \ {d \over
dr} f_I(r)
\end{equation}
and
\begin{equation}
U_S(r)=2\ \Big( {\hbar \over m_\pi c}\Big)^2 \ V_S
\ {1\over r}\ {d \over dr}f_S(r)
\end{equation}
are the central and spin-orbit part of the potential, respectively, and
$U_C(r)$ is the proton-nucleus Coulomb potential. The
Fermi functions $f_i$  are given by
\begin{equation}
f_i(r)=\Bigl\{ 1+\exp[(r-R_i)/a_i]\Big\}^{-1} \ ,
\end{equation}
with $R_i=r_iA^{1/3}$.

In the eikonal approximation, the proton-nucleus
elastic scattering cross section is given by \cite{Gl59}
\begin{equation}
{d \sigma_{el} \over d \Omega}= \Big| F(\theta) \Big|^2
+\Big| G(\theta) \Big|^2 \ ,
\end{equation}
where
\begin{equation}
F(\theta)=f_C(\theta)+ik\ \int_0^\infty db \ b\ J_0(qb)
\ \exp \Big[ i\chi_C(b)\Big]\ \bigg\{ 1- \exp\Big[ i\chi_0(b)\Big] \
\cos\Big[kb\
\chi_S(b)\Big]\bigg\}
\end{equation}
and
\begin{equation}
G(\theta)=ik\ \int_0^\infty db \ b\ J_1(qb)
\ \exp\Big[ i\chi_C(b)+i\chi_0(b)\Big] \ \sin\Big[kb\
\chi_S(b)\Big] \ .
\end{equation}
In the equation above $q=2k\sin(\theta/2)$, where $\theta$ is
the scattering angle, $J_0$ ($J_1$) is the zero (first) order
Bessel function.
The eikonal phase $\chi_{0(S)}$ is given by
\begin{equation}
\chi_{0(S)}({\bf b}) = -{1 \over \hbar v} \
\int_{-\infty}^\infty U_{0(S)} ({\bf b}, \ z)\ dz
\ .
\end{equation}
For the Coulomb eikonal phase we use
the approximation, valid for a point nucleus,
\begin{equation}
\chi_C(b)=\eta \ \ln (kb)\qquad {\rm with} \quad \eta=
{2 Z_1 Z_2 e^2\over \hbar v} \ ,
\end{equation}
where $Z_1$ and $Z_2$ are the proton ($Z_1=1$) and the nuclear charges,
respectively.
The Coulomb phase will be changed
for a finite charge distribution
of the nucleus \cite{Fa72}. For example, assuming
a uniform charge distribution with radius $R$ the Coulomb
phase becomes
\begin{eqnarray}
\chi_C(b)&=&\eta \ \bigg\{ \Theta (b-R) \ \ln (kb)
+\Theta (R-b)\Big[ \ln (kR)+\ln (1+\sqrt{1-b^2/R^2})\\ \nonumber
&-&\sqrt{1-b^2/R^2}-{1\over 3} (1-b^2/R^2)^{3/2}\Big] \bigg\} \ ,
\end{eqnarray}
where $\Theta$ is the step function.
This expression is finite for $b=0$, contrary to eq. (9). If one
assumes a gaussian distribution of charge
with radius $R$, appropriate for light
nuclei, the Coulomb phase becomes
\begin{equation}
\chi_C(b)=\eta \ \Big\{ \ln (kb) + {1\over 2} E_1(b^2/R^2)\Big\}
\ ,
\end{equation}
where the error function $E_1$ is defined as
\begin{equation}
E_1(x)=\int_x^\infty {e^{-t}\over t} \ dt  \ .
\end{equation}
This phase also converges, as $b\rightarrow 0$.

We have noticed in our numerical calculations that the above corrections
to the approximation (9)
do not modify the calculated cross sections appreciably.
Moreover, one can show that in the eikonal approximation, the
phase defined in eq. (9) yields exactly the Coulomb scattering amplitude
\begin{equation}
f_C(\theta)={Z_1Z_2e^2\over 2 \mu v^2 \ \sin^2 (\theta/2)}
\ \exp\Big\{ - i \eta \ \ln \Big[
\sin^2(\theta/2) \Big] +i\pi+2 i\phi_0\Big\}
\end{equation}
where $\phi_0=arg\Gamma (1+i\eta/2)$.
This is convenient for
the numerical calculations since eq. (6) is written with the
separated contribution of the Coulomb scattering amplitude.
Then, the remaining integral (the second term on the r.h.s. of eq. (6))
converges rapidly for the scattering at forward angles.
If we use eqs. (10)and (11) the amplitude $F(\theta)$
cannot be separated as in eq. (6)
and the integral in $b$ will converge very slowly.

A more important correction, due to the Coulomb deflection of the
trajectory, amounts to calculate all elastic and inelastic integrals
(to be discussed in section 4) replacing the asymptotic impact
parameter $b$ by the distance of closest approach in Rutherford orbits,
i.e.,
\begin{equation} kb'=\eta+\sqrt{\eta^2+k^2b^2} \ .
\end{equation}
As shown by Vitturi and Zardi
\cite{VZ87}
this correction leads to a considerable
improvement of the eikonal amplitudes for the scattering of heavy systems.

For nucleus-nucleus collisions the spin-orbit interaction, $U_S(r)$,
and the surface-term of the imaginary potential
(last term of eq. 2) are usually neglected.
Whereas these terms are relevant for proton-nucleus scattering, they
play no important role in  nucleus-nucleus collisions.
Thus, for nucleus-nucleus collisions, the scattering
amplitude is given by
\begin{equation}
{d \sigma_{el} \over d \Omega}= \Big| F(\theta) \Big|^2\ ,
\end{equation}
where
\begin{equation}
F(\theta)=f_C(\theta)+ik\ \int_0^\infty db \ b\ J_0(qb)
\ \exp\Big[ i\chi_C(b)\Big] \Big\{ 1- \exp\Big[ i\chi_0(b)
\Big] \Big\} \ .
\end{equation}
Obviously, these equations can be obtained from eqs. (5-7)
by setting $\chi_S=0$.

A common way to relate the nuclear optical potential to
the ground-state densities is to use the ``t$\rho\rho$"
approximation. This approximation has been extensively
discussed in the literature
\cite{HRB91,LVZ88}. In its simplest version, neglecting the spin-orbit
and surface terms,
the
optical potential for proton-nucleus collisions is given by
\begin{equation}
U_0({\bf r})=<t_{pn}>\rho_n({\bf r})+<t_{pp}>\rho_p ({\bf r})
\end{equation}
where $\rho_n$ ($\rho_p$) are the neutron (proton)
ground state densities and $<t_{pi}>$ is the
(isospin averaged) transition
matrix element for nucleon-nucleon scattering at forward
directions,
\begin{equation}
t_{pi}({\bf q}=0)
=-(2\pi\hbar^2/\mu)\ f_{pi}(
{\bf q}=0)
=-{\hbar v\over 2} \ \sigma_{pi} \ (\xi_{pi}+i)
\end{equation}
where $\sigma_{pi}$ is the free proton-nucleon cross
section and $\xi_{pi}$ is the ratio between the imaginary
and the real part of the proton-nucleon scattering amplitude.
The basic assumption here is that the scattering is given
solely in terms of the forward proton-nucleon scattering
amplitude and the local one-body density \cite{HRB91}.

For nucleus-nucleus collisions, we will use the same method which
leads to an optical potential of the form
\begin{equation}
U_0({\bf R})=\int <t_{NN}({\bf q}=0)> \ \rho_1 ({\bf R-r'}) \ \rho_2
({\bf r'}) \ d^3r' \ ,
\end{equation}
where ${\bf R}$ is the distance between the
center-of-mass of the nuclei. We use the
isospin average $<t_{NN}>=(t_{pp}+t_{pn})/2$.

Without much computational effort the formula (19) is
improved to account for the scattering angle dependence of the
nucleon-nucleon amplitudes. A good parametrization \cite{Ra79} for the
nucleon-nucleon scattering amplitude is given by
$f_{NN}({\bf q})=
(k_{NN}/4\pi) \ \sigma_{NN} \ (i+\alpha_{NN})\ e^{-\xi_{NN}q^2} $.
The nuclear scattering phase then becomes
\cite{Gl59}
\begin{equation}
\chi_0(b)=\int \int d {\bf r} \ d{\bf r}' \ \rho_1({\bf r}) \
\gamma_{NN}(|{\bf b-s-s'}|) \ \rho_2({\bf r}')
\end{equation}
where the profile function $\gamma_{NN}(b)$ is defined in terms
of the two-dimensional Fourier transform of the elementary
scattering amplitude
\begin{equation}
\gamma_{NN} (b)= {1 \over 2\pi i k_{NN}} \ \int \exp\Big[-i{\bf q.b}\Big]
\ f_{NN}({\bf q}) \ d{\bf q}
\ ,
\end{equation}
and ${\bf s, \ s'}$ are the projections of the coordinate vectors
${\bf r}$, ${\bf r}'$
of the nuclear densities on the plane perpendicular to the $z$-axis.
For spherically symmetric ground-state densities eq. (20)
simplifies to the expression
\begin{equation}
\chi_N(b)=\int_0^\infty dq \ q \  \hat \rho_1(q) \
f_{NN}(q) \ \hat \rho_2(q) \ J_0(qb) \ ,
\end{equation}
where $\rho_i(q)$ are the Fourier transforms of the ground state
densities.

In the following we will
use the eq. (17) in order to calculate
the proton-nucleus eikonal phase-shifts. For nucleus-nucleus
collisions we will use the formula (22), since we assume the
spherical symmetry for the ground state distributions.
We will now
describe how we calculate the ground state densities for
exotic nuclei.

\bigskip\bigskip

\section{Ground State Densities}

We calculate the density distributions of halo nuclei based on the Hartree-
Fock (H-F) approximation with Skyrme interaction.  The Skyrme interaction is
known to describe successfully the ground state properties
(the binding energies, the rms radii and the  charge
distributions) of many nuclei
in a broad region of the mass table \cite{VB72}.

It was pointed out that the separation energy of loosely bound neutrons
plays an important role to study halo nuclei \cite{HJ87,BBS89,Bro91}.
One needs an accuracy of about a few tens of
keV to take into account the effect
of the separation energy on the radii and the density distributions of
halo nuclei, since they have extremely small
separation energies as is shown in table 1.
On the other hand,
the H-F theory cannot provide the prediction
for the separation energies within the accuracy of a few tens of keV
\cite{Mah85}. It is known that
higher order effects
beyond the mean field approximation are necessary to describe
more precisely the single-particle energies
near the Fermi surface\cite{HS76}.
So far, several theoretical attempts\cite{BG80}
have been made in this direction,
but the results do not satisfy the accuracy
which is required for the study of halo nuclei.
We take the following method\cite{BBS89,Bro91,Sag92}
to improve the calculated separation energies rather than evaluating directly
the higher order effects; the last neutron configuration is treated
differently from the other orbits in the H-F potential in order to
reproduce properly the neutron separation energy of the nucleus.

In the above procedure,
the single-particle energy of the last orbit is adjusted to be
the same as the empirical single-neutron separation energy in the case of odd-N
system, while the empirical two-neutron separation energy is adopted
for the last neutron orbit in the even-N nucleus.
Although it is not obvious how much the two-body correlation between
the halo neutrons affects the single-particle wave function,
this choice for the even-N nucleus was pointed out reasonable
to study the Coulomb dissociation cross sections $^{11}$Li
\cite{Iek93}. The soft dipole excitation in $^{11}$Be \cite{Nak94} is
also described well by this wave function.

The H-F equation for the Skyrme interaction can be written  as
\begin{equation}
\Bigl[-\nabla\frac{\hbar^{2}}{2m^{*}(r)}\nabla + V(r) \Bigr]
   \psi_{\alpha}(r)= \epsilon_{\alpha} \psi_{\alpha}(r)
\end{equation}
where m$^{*}$(r) is the effective mass.
The potential  V(r) has a central, a spin-orbit and a Coulomb term,
\begin{equation}
V(r)= V_{central}+V_{spin-orbit}+V_{Coulomb}.
\end{equation}
This H-F potential can be expressed analytically in terms of the parameters
of the Skyrme interaction\cite{VB72}.
The central potential in eq. (2)
is multiplied by a constant normalization factor ${\sl f}$ only for the last
neutron configuration:
\begin{eqnarray}
V_{central}(r) = fV_{H-F}(r), \hspace*{0.5cm}
       \left  \{ \begin{array}{ll}
             f \ne 1 & {\mbox for\, last\, neutron\, configuration}  \\
                  f  =  1 & {\mbox otherwise}   \end{array} \right.  .
\end{eqnarray}
  Numerical calculations are performed with the parameter set SGII\cite{GS81}
which  gives satisfactory results for
charge distributions of many nuclei, and also
for the systematics of the nuclear radii in comparison with experimental
data\cite{Bro84}.

The empirical separation energies of light neutron-rich nuclei
are tabulated in table 1.  Nuclei with negative values in table 1
are unstable against neutron decay.
We can see that the separation energies
are extremely small in the cases of $^{6}$He, $^{11}$Li,
$^{11}$Be and  $^{14}$Be which are known as halo nuclei.
     Calculated radii are also tabulated in table 2 together with empirical
data.

\section{Results for Elastic Scattering}

We apply the formalism presented in section 2 to the
elastic scattering of stable nuclei in order to study
the validity of the method. We will
concentrate here in nuclei with spherical symmetry.
The ground state densities of stable nuclei
used here are parameterized as gaussian (G) and
modified Fermi (MF) densities, as shown in table 3.
The nucleon-nucleon
cross sections used as input to construct the optical
potentials in the ``$t\rho\rho$" approximation are shown in
table 3. A linear interpolation is done to find the appropriate
set of parameters at energies between those shown in the table.
For energies lower than 94 MeV/nucleon we use $\xi_{NN}=0.5$ $fm^2$.
As one can deduce from eq. (22) the $q$-dependence of $f_{NN}$ is
very important with the parameters $\xi_{NN}$, as given in table 4
since the integrand in eq. (22) is only relevant for
$q < 1/b\ll 1/\sqrt{\xi_{NN}}$.
The parameters for the densities of stable
nuclei used are shown in table 3.

In fig. 1
we show the experimental data from ref. \cite{Ba88} for the
elastic scattering of $^{17}O+^{208}Pb$ at 84 MeV/nucl.
The curve is calculated using the nuclear phase-shift
constructed as in eq. (22) and the scattering amplitude as
in eq. (16). We observe that the agreement with the data is
extremely good. The rainbow scattering at $\theta\sim2.5^\circ$
is also very well reproduced. However, one should be cautious
with such an example since in this case the scattering amplitude
is dominated by a sharp
transition from no-absorption to
strong absorption as the scattering angle increases. Basically,
the optical potential has to have the feature of a
large absorption
at small distances and a small diffuseness.
This behavior arises naturally within the framework of the ``$t\rho\rho$"
approximation for heavy systems.

Light systems are  more ``transparent" and the collision at small
distances (large scattering angle) are more sensitive to the
details of the optical potential. To show this we plot in
fig. 2 the elastic scattering of $\alpha+\alpha$ at $E_{lab}=
2.57$ GeV as a function of the invariant momentum transfer
$t=-(p_1+p_2)^2$.
The data are from ref. \cite{Be80}.  The calculated curve agrees reasonably
well with the data at forward angles (small $t$) but deviates
appreciably from it at larger angles.
The
forward scattering is dominated by peripheral
collisions for which  multiple scattering is not relevant. Thus,
we expect that the ``$t\rho\rho$'' approximation works well at forward
angles.
On the other hand, it has been shown
that multiple collisions are very important for large scattering
angles \cite{CM69}, so that the ``t$\rho\rho$ approximation
fails to explain the data.

We plot in fig. 3 the elastic scattering  of $^{12}C+^{12}C$
at $E_{lab}=85$ MeV/nucl to study the model further.
The data are from ref. \cite{Bu82}. The solid
curve is obtained by using the eikonal approximation (eqs. 16
and 22) with a set of Woods-Saxon potentials which was constructed
to give the smaller
$\chi$-square fit
to the data with a DWBA calculation using the
code PTOLEMY \cite{MP78}. This Woods-Saxon potential reproduces the data
perfectly \cite{Bu82}.
The parameters obtained by this fit are
\begin{eqnarray}
V_0&=&-120 \ {\rm MeV}, \qquad R_v=1.72 \ {\rm fm}, \quad {\rm and}
\qquad a_v=0.83 \ {\rm fm}\\ \nonumber
W_0&=&-46.8 \ {\rm MeV}, \qquad R_w=2.2 \ {\rm fm}, \quad {\rm and}
\qquad a_v=0.86 \ {\rm fm} \ .
\end{eqnarray}
The dashed curve is obtained with the ``$t\rho\rho$" approximation which
shows again the mismatch with the data at large scattering angles.
We see that the ``t$\rho\rho$ approximation gives a reasonable
description of the elastic scattering only at forward angles. It is known
that large scattering angles are affected by corrections due to
real, and virtual (polarization) nuclear excitation \cite{Tak93}.
This means that a simple and unambiguous relationship between
the elastic scattering data
and the nuclear ground-state densities does not exist at large scattering
angles.
We conclude that
the disagreement between the experimental data
and the method used here, for light
systems and
large scattering angles, is not a deficiency of the eikonal
approximation, but of the construction of the optical potential
(i.e. on the assumption of the validity of the ``t$\rho\rho$''
approximation).

The corrections due to multiple collisions using the Glauber
multiple scattering series is rather cumbersome \cite{CM69}
and a direct
connection between the ground state densities of the nuclei
and the scattering data is lost among a large  variety of
approximations. We feel that, due to its simplicity, the
``$t\rho\rho$" approximation is very appealing within  its limitations.
We therefore will use this approximation since
the complications arising from a treatment of multiple scattering,
and of other effects, undermines our effort to extract information about
the ground state densities of the nuclei from the elastic scattering.
We will focus the following study mainly in the forward scattering
region because of the limited validity of the model.

Fig. 4(a) shows the elastic scattering of proton on $^9Li$ at
$E_{lab}=60$ MeV. The data are from ref. \cite{Mo92}.
The solid curve is
obtained by using eqs. (5-7) with  the parameters
shown in table II  of ref. \cite{Mo92} which
was obtained by a $\chi$-square fitting to the data with a DWBA
calculation.
The dashed curve is obtained by using the optical potential
constructed as in eqs. (17-19) together with the Hartree-Fock $^9Li$
ground-state density.
In this case the surface and
spin-orbit interaction are absent. The dashed curve
clearly misses the
experimental data.
The absorption is greater than expected by the ``t$\rho\rho$" method so
that
this difference cannot be
ascribed only to the absence of the spin-orbit interaction.
Quite a different scenario is presented in figure 4(b)
where we plot the $p+^{11}Li$ scattering data
at 62 MeV/nucleon from
ref. \cite{Mo92}. The solid curve is calculated with the optical potential
parameters of table II (set B)  of ref. \cite{Mo92} for eqs. (1-3),
which were chosen
so as to fit the experimental data.
The dashed curve is obtained by
using the eqs. (17-19) and the $^{11}Li$ Hartree-Fock
ground state density, calculated as explained in section 2.
The agreement with the data is quite good at forward angles,
in contrast to the previous case.

One might think that the good quantitative
description obtained in figure (4b) is accidental, in view of the
previous discussions. In order to clarify further the validity
of the model,
in fig.  5 we show the elastic scattering data of $p+^8He$ at
72 MeV/nucleon together with a calculation using the optical
potential constructed as in eqs. (17-19) by the solid curve.
The ground state density
of $^8He$ was calculated as explained in section 3.
Although the data uncertainties at $\theta > 40^\circ$
do not allow for a test of the theory, the agreement
with the data is almost perfect
at forward angles.
Such an agreement is very
encouraging since
the simple ``$t\rho\rho$" approach is of very
useful predictive power and  it can be used to plan  future
experiments on radioactive beams scattering of protons.
Also shown in fig. 5, the dashed curve is the scattering
cross section obtained by using the $^6He$ ground state density.
Two important features are seen. Firstly, the strength of the
strong absorption in the $p+^8He$ system is appreciably larger
than in the $p+^6He$ system. Secondly, the diffraction minimum is
at a larger angle for $^6He$ targets than for the $^8He$ ones,
revealing a larger absorption radius of $^8He$.

The inclusion of the surface and spin-orbit terms of eq. (1) in
a  microscopic Glauber approach, as in
eq. (17), to elastic scattering is rather
complicated (see discussion in ref. \cite{HRB91}). The absence of these
terms in our approach may be considered as one of the  main
reasons for the
discrepancies between the experimental data and the calculations. This
problem should not occur for nucleus-nucleus collisions, since
the folding of densities ``smear out" the
surface and spin-orbit terms of the nucleon-nucleus
optical potential.

Data on elastic scattering of halo nuclei has been taken
during the last few years. Due to the poor energy resolution,
the data are contaminated with inelastic scattering. In fig.
6, 7 and 8
we plot the data on quasi-elastic scattering of $^{11}Li+^{12}C$
at $E_{lab}=637$ MeV, of $^{12}Be+^{12}C$ at $E_{lab}=679$ MeV,
and of $^{14}Be+^{12}C$ at $E_{lab}=796$ MeV, respectively.
The dashed curves are calculated by  using the
prescription given by eqs. (15-16, 22) and the Hartree-Fock
densities calculated as explained in section 2.
The data are from refs. \cite{Ko92,Za94}.
While the
agreement is quite reasonable for the $^{12}Be+^{12}C$
and $^{14}Be+^{12}C$ data,
it  does not work so well for the $^{11}Li+^{12}C$ data.
The solid curves are calculated by including the inelastic
excitation of the  $2^+$, and the $3^-$ states in $^{12}C$ which
cannot be separated from the experimental data.
In the next section we describe how to calculate the inelastic
scattering excitation cross sections.
We use the deformed potential model
for the excitation of $^{12}C$, with deformation parameters
$\beta_2=0.59$ and $\beta_3=0.40$ for the $2^+$ and the $3^-$
states, respectively. These are the same values used in refs.
\cite{Ko92,Za94}.
We observe that, with
the inclusion of inelastic scattering, the theoretical
results for the $^{11}Li+^{12}C$ collision is remarkably
good, but the ones for $^{12}Be+^{12}C$ and for $^{14}Be+^{12}C$
are not, yielding too large cross sections at large angles.
One possible reason for such a discrepancy is the
the absence of
dynamical polarization in our approach. The virtual excitations
during the collision time is shown to affect the elastic scattering
of halo neutron appreciably and has been
studied by several authors \cite{BCH93}. An inclusion of such effects
is beyond our model of using a simplified approach to study
the ground-state properties of the halo nuclei.
Another possibility is that
the folding prescription to determine the optical
potential yields
larger cross sections than the experimental ones,
as we observed in fig. 3.

We conclude that the simple ``t$\rho$$\rho$" folding procedure
is an useful method to describe quantitatively the elastic
scattering data of radioactive nuclei in most cases.
We should also notice that the inelastic scattering makes it
difficult to obtain good agreement with the data, especially at
larger angles.
However, a test of the ground-state
densities of such nuclei is possible by comparing the data with the
theoretical calculations at forward angles.
At large angles, especially
with the presence of inelastic scattering, the results are however
sometimes misleading. More detailed microscopic calculations would be
necessary to obtain better agreement. In the opposite extreme,
one may resort
to the use of optical potential parameters to fit the data.
In both cases, a link of the results to microscopic
features of the
nuclei is very difficult to achieve.
\bigskip\bigskip

\section{\bf Inelastic Scattering}
\subsection{\bf Nuclear Excitation}
\medskip

We can use the Distorted Wave Approximation
(DWBA) for the inelastic amplitude,
assuming that a residual interaction $U$ between the projectile and
the target exists and is weak . The cross section
for the excitation of a vibrational mode $(\lambda \mu)$ with
energy $\hbar \omega_\lambda$ is given by
\begin{equation}
{d \sigma_{\lambda\mu} \over d \Omega} =
{k_\lambda \over k_0} \ |f_{\lambda \mu} (\theta) |^2 =
\Big( {M \over 2 \pi \hbar^2} \Big)^2 \ {k_\lambda \over k_0} \
\Big| <\Psi_{\lambda \mu} \ \phi^{(-)}_{k_\lambda} |U|
\Psi_0 \ \phi^{(+)}_{k_0}>
\Big|^2
\end{equation}
where $k_\lambda$ is defined as
\begin{equation}
E_{k_\lambda}={\hbar^2 k_\lambda^2 \over 2 M}= {\hbar^2 k_0^2 \over 2 M}
-\hbar \omega_\lambda \ .
\end{equation}
In eq. (28), $M$ is equal to the reduced mass of the system and $k_0$
is equal their relative
momentum. The wavefunctions $\phi$ and $\Psi$ describe the relative
motion and the internal states, respectively.

In the particle-vibration-coupling model  the
transition matrix elements are given by
\begin{equation}
M_{\lambda \mu} ({\bf r})
\equiv <\Psi_{\lambda\mu}|U|\Psi_0>=
{\delta_\lambda \over \sqrt{2\lambda +1}}
\ Y_{\lambda \mu} (\hat {\bf r})
\ U_{\lambda} (r)
\end{equation}
where $\delta_\lambda=\beta_\lambda R$
is the vibrational amplitude, or {\it
deformation length},
$R$ is the nuclear radius, and $U_{\lambda}
(r)$ is the transition potential. We will consider only low multipolarities,
$l\le 2$ in the following.

The deformation length $\delta_\lambda$ can be directly related to the
reduced matrix elements for electromagnetic transitions. Using well-known
sum-rules for these matrix elements one finds a relation between the
deformation length,
and the nuclear sizes and the excitation energies.
For isoscalar excitations one obtains \cite{Sa87}
\begin{equation}
\delta_0^2= 2 \pi \ {\hbar^2 \over m_N} \
{<r^2> \over A E_x} \ , \
\delta_{\lambda \geq 2}^2 = {2 \pi \over 3} \ {\hbar^2 \over m_N} \
\lambda \ (2\lambda +1) \ {1\over A E_x}
\end{equation}
where $A$ is the atomic
number, $<r^2>$ is the r.m.s. radius of the nucleus,
and $E_x$ is the excitation energy.

The transition potentials for isoscalar excitations are
\begin{equation}
U_0 (r) = 3 U_{opt} (r) + r {d U_{opt} (r) \over dr}
\ ,
\end{equation}
for monopole, and
\begin{equation}
U_2 (r)= {dU_{opt} (r) \over dr} \ ,
\end{equation}
for quadrupole modes.

For dipole isovector excitations, the deformation length is given by
\begin{equation}
\delta_1= \pi \ {\hbar^2 \over m_N}
\ {A \over NZ} \ {1\over E_x}\ ,
\end{equation}
where $Z$ ($N$) the charge (neutron)  number. The transition potential
in this case is \cite{Sa87}
\begin{equation}
U_1(r)=\gamma \ \Big( {N-Z \over A} \Big) \
\Big( {dU_{opt} \over dr} + {1\over 3} \ R_0 \ {d^2 U_{opt}
\over dr^2} \Big)
\ ,
\end{equation}
where the factor
$\gamma$ depends on the difference between the proton and the neutron
matter radii  as
\begin{equation}
\gamma {2(N-Z)\over 3A} = {R_n-R_p \over {1\over 2} \ (R_n+R_p)}
= {\Delta R_{np} \over R_0}
\ .
\end{equation}
Thus, the strength of isovector excitations increases with
the difference
between the neutron and the proton matter radii.
This difference is accentuated for neutron-rich
nuclei
so that the isovector
dipole excitations should be a good test for the quantity $\Delta R_{np}$
which becomes very large for neutron-rich unstable nuclei. One
can
generalize this equation to higher isovector multipole
excitations $\lambda \ge 2$ by the substitution
\begin{equation}
{N-Z\over A} \ \ \longrightarrow \ \ Q_\lambda^{(n)} +
Q_\lambda^{(p)}=Z\Big(-{1\over A}\Big)^\lambda +
\Big[ \Big(1-{1\over A}\Big)^\lambda +(-1)^\lambda {(Z-1) \over A^\lambda}
\Big]         \ ,
\end{equation}
where $Q_\lambda^{(n,p)}$ are the effective charges (in units of $e$) of
the neutron and the proton, respectively.

An useful approximation, valid for $\hbar \omega_\lambda \ll E_{k_\lambda}$,
is
\begin{equation}
k_\lambda \simeq k_0 \ \Big( 1 - {M\omega_\lambda \over \hbar k_0^2}
\Big) = k_0+ {\omega_\lambda \over v}
\ .
\end{equation}
Further, by using the definition (27) and the eikonal approximation
\begin{equation}
\phi^{(-)*}_{k_\lambda}\ \phi_{k_0}^{(+)} \simeq
\exp \Big\{ i{\bf q.r}+i\chi(b)\Big\}
\ ,
\end{equation}
we get
\begin{eqnarray}
f_{\lambda\mu} (\theta)&=&{M\over 2 \pi \hbar^2} \
{1\over \sqrt{\lambda+1}} \ \delta_\lambda \ i^\mu
\ \sqrt{\pi (2\lambda+1)} \ \sqrt{(\lambda -\mu)!\over
(\lambda + \mu)!} \\  \nonumber
&\times& \int_0^\infty db \ b\ J_\mu (q_t b) \ e^{i\chi (b)}
\int_{-\infty}^\infty dz\
P_{\lambda\mu} \Big( {z \over \sqrt{b^2+z^2}}\Big) \ U_\lambda (b,z) \
e^{i\omega_\lambda z/v} \ ,
\end{eqnarray}
where $J_\mu$ and $P_{\lambda\mu}$ are the Bessel function and the
Legendre polynomials, respectively, and
$q_t=2\sqrt{k_0k_\lambda} \sin (\theta / 2)$.

As seen in eq. (39), one needs to calculate two
simple integrals to compute the inelastic scattering
at intermediate energies
with the deformed potential model.
The scattering amplitudes will depend on the optical
potential parameters and on the deformation length $\delta_\lambda$.
The deformed potential model is based on the assumption of a transition
density peaked at the surface. Although this assumption is
reasonable for the
excitation of heavy nuclei ( e.g., $^{40}Ca$, $^{208}Pb$),
it is rather crude for light nuclei, especially when the transition density
extends radially beyond the nuclear size. This is the case for the
soft multipole excitations, for which the transition
densities have very long tails.

A more convenient way to describe the inelastic scattering
of neutron-rich nuclei is to use the
folding approximation. The assumption of a transition
density peaked at the surface of the nucleus is not necessary. In this
model the matrix element on the right-hand-side of eq. (27) is
\begin{equation}
T_{\lambda\mu}=\int d^3R \int d^3r \
\phi^{(-)*}_{{\bf k}_\lambda}({\bf R}) \
U_{int} \Big( |{\bf R-r}|\Big) \
\delta\rho_{\lambda\mu}({\bf r}) \ \phi_{k_0}^{(+)}({\bf R})
\ ,
\end{equation}
where $\delta \rho_{\lambda\mu} =\Psi^*_{\lambda\mu}\Psi_0$ is the
transition density. $U_{int}\Big(|{\bf R-r}|\Big)$ is the potential
between each nucleon of the target and the projectile nucleus. Thus,
the transition from the ground state to the
excited state is directly calculated
from the target nucleon-projectile interaction.

For light targets, a gaussian parametrization of the
(target nucleon)-projectile potential
is adequate and yields simple formulas.
This can be shown by means of
the expansion
\begin{eqnarray}
U_{int}\Big(|{\bf R-r}|\Big)&=&(v_0+iw_0)\ e^{-({\bf R-r})^2/a^2} \\ \nonumber
&=&4\pi (v_0+iw_0) \ e^{-(R^2+r^2)/a^2} \ \sum_{\lambda\mu}
i^\lambda \ j_\lambda (2i{rR\over a^2}) \ Y_{\lambda\mu}
(\hat {\bf R})\ \ Y_{\lambda\mu}^*(\hat {\bf r})
\ ,
\end{eqnarray}
where $j_\lambda (ix)$ are the spherical Bessel functions calculated
for imaginary arguments.

Using this result in the eq. (27) and the definition
$\delta \rho({\bf r})= \delta \rho_\lambda
(r)\ Y_{\lambda\mu}(\hat {\bf r})$
we get (using ${\bf R}=({\bf b}$, $Z)$)
\begin{eqnarray}
T_{\lambda\mu}&=&4\pi^{3/2}\ (v_0+iw_0) \ \sum_{\lambda\mu}
i^{\lambda\mu} \sqrt{(2\lambda+1) \ (\lambda -\mu)! \over
(\lambda+\mu)!} \\  \nonumber
& \times & \int_0^\infty db \ b\ J_\mu (q_tb) \
e^{i \chi(b)} \ {\cal O}_{\lambda\mu} (b)
\,
\end{eqnarray}
where
\begin{equation}
{\cal O}_{\lambda\mu}(b)=\int_0^\infty dr \ r^2 \delta
\rho_{\lambda\mu} (r) \ F_{\lambda\mu} (r,b) \ ,
\end{equation}
with
\begin{equation}
F_{\lambda\mu} (r,b)=\int_{-\infty}^\infty dZ \ \exp \Big(
-{{R^2+r^2}\over a^2}\Big) \ j_\lambda \Big({2 i r R \over a^2}\Big)
\ P_{\lambda\mu} \Big( {Z \over R}\Big) \ \exp \Big(i{\omega_\lambda Z
\over v}\Big)
\ .
\end{equation}

A link between the deformed potential model and the folding model
is obtained by using the approximation
\begin{equation}
\delta \rho_\lambda = -
\left\{ \begin{array}{ll}
           \delta_\lambda \ d\rho_0/dr/\sqrt{2\lambda+1},
           & \mbox{for $\lambda \ge 1$; }\\
    \delta_0\ \Big(3 \rho_0+ r d\rho_0/
             dr\Big),
           & \mbox{for $\lambda=0$;}
\end{array}
\right.
\end{equation}
As in the deformed potential model, the scattering amplitude is determined
by the optical potential parameters and the deformation length
$\delta_\lambda$.

Another common approximation for $\delta\rho_\lambda$ is
provided by the Tassie model \cite{Ta56}
which gives
\begin{equation}
\delta \rho_\lambda (r)= - {\delta_\lambda \over \sqrt{2\lambda+1}}
\ \Big( {r \over R_0} \Big)^{\lambda -1} \ {d \rho_0 \over dr}
\ ,  \ \ \ \ {\rm for} \ \ \ \lambda\ge 1 \ .
\end{equation}
For $\lambda=0$, one
uses eq. (45). In general, both models yield analogous transition densities
for heavy nuclei and low collective states.

These approximations assume that
the transition density is peaked at the nuclear surface. As we will
show later, this is a bad approximation for neutron rich nuclei. It is
more convenient to use directly the transition density calculated from
microscopic models and inserted
into eqs. (42) and (43) to obtain the scattering
amplitude.
\bigskip\bigskip
\subsection{\bf Coulomb excitation}
The subject of Coulomb excitation for heavy ion collisions at
intermediate energies,
has been  discussed in ref. \cite{BN93}, including effects of retardation
and strong absorption effects.
The Coulomb excitation
amplitude
for a given multipolarity $E\lambda$ is given by
\begin{equation}
f_{E1, \mu}(E_x, \theta)=i {\sqrt{8\pi} \over 3} \
{Z_T e M_{aA} \over \hbar^2}
\ \Bigl( {E_x \over \hbar v_a} \Bigr) \
\Bigl[ B(E1, \ E_x) \Bigr]^{1/2} \
\
\left\{ \begin{array}{ll}
           \Lambda_{\pm 1} (E_x, \ \theta),  & \mbox{for $\mu=\pm 1$; }\\
           \Lambda_0 (E_x, \ \theta),    & \mbox{for  $\mu=0$}
\end{array}
\right.
\label{atop1}
\end{equation}
for E1 excitations, and
\begin{eqnarray}
f_{E2, \mu}(E_x, \theta)&=& - \ {2\over 5} \ {\sqrt{\pi} \over 6} \
{Z_T e M_{aA} \over \hbar^2}
\ \Bigl( {E_x \over \hbar v_a} \Bigr)^2 \
\Bigl[ B(E2, \ E_x) \Bigr]^{1/2} \
\nonumber \\
\nonumber \\
&\times& \left\{ \begin{array}{ll}
    \Lambda_{\pm 2} (E_x, \ \theta)/\gamma, & \mbox{for $\mu=\pm 2$; }\\
   -(2-v^2/c^2) \ \Lambda_{\pm 1} (E_x, \ \theta), & \mbox{for  $\mu=\pm 1$;
}\\
   \Lambda_0 (E_x \ \theta),  & \mbox{for  $\mu=0$}
\end{array}
\right.
\label{atop2}
\end{eqnarray}
for E2 excitations.

In the above equations $\gamma=(1-v_a^2/c^2)^{-1/2}$, and
$\mu$ is the azimuthal component of the
transferred angular momentum. The functions $\Lambda_\mu$ are given
by
\begin{equation}
\Lambda_\mu (E_x, \ \theta)=\int_0^\infty db \ b \
J_\mu(qb) \ K_\mu\Bigl({E_x b \over \gamma \hbar v_a}\Bigr) \
\exp\Bigl[ i \chi(b) \Bigr]
\end{equation}
where $q=2k \sin(\theta/2)$, $k$ is  the
c.m. momentum of $a+A$,
and $J_\mu$ ($K_\mu$) are Bessel (modified)
functions of order $\mu$. $\chi(b)=\chi_N(b)+\chi_C(b)$
is obtained from eqs. (8) and (9).

Assuming that an isolated state is excited, and that it exhausts fully
the sum rules, one gets ($B(E\lambda)\equiv B(E\lambda, E_x)$)
\begin{equation}
B(E1)= {3 \over 4\pi} \ {\hbar^2 \over 2 m_N} \
{NZ\over AE_x} \ e^2
\ ,
\end{equation}
and
\begin{equation}
B(E2)= {\hbar^2\over m_N} \ {3R^2\over 4\pi E_x} \ e^2
\times \left\{ \begin{array}{ll}
    Z^2/A, & \mbox{ for isoscalar excitations; }\\
   NZ/A, & \mbox{for isovector excitations; }
\end{array}
\right.
\end{equation}

The electromagnetic transition operators have a smooth dependence on
the spatial coordinates (${\cal O}(E1) \sim {\bf r}$ and
${\cal O}(E2) \sim r^2$). Thus, the electromagnetic excitation has a very
weak dependence on the spatial form of the transition density
and it does not
seem to be a  direct probe of the halo properties.
However, the reduced matrix elements
for halo nuclei are expected to be enhanced due to unique properties of the
halo wavefunctions. For example, if the binding of the valence
neutrons is very weak, as in the case of $^{11}Li$, one expects a huge
enhancement of the $B(E1)$-strength at low energies (soft E1-modes).
This can be thought as due to the threshold effect of extended
halo neutrons with respect to the
core, which leads to large B(E1)-values.
\bigskip\bigskip

\section{\bf Results for Inelastic Scattering}
\bigskip
We first apply the formulation of the previous section to the inelastic
scattering of stable nuclei.
In fig. 9 we plot the inelastic scattering of 84 MeV/nucleon $^{17}O$
projectiles on lead. The excitation of the isoscalar
giant quadrupole resonance in
lead ($E_x= 10.9$ MeV) is considered.
Data points are from ref. \cite{Ba88}. The curves are calculated
by using the deformed potential model.  The optical potential was calculated
by using the folding procedure described in section 2. The deformation
parameter $\beta_2=0.63$ was used. This corresponds to 100\% of exhaustion
of the sum rule.  The dashed curve is the contribution of the
Coulomb excitation with  $B(E2)=0.73$ $e^2$ $b^2$, also corresponding to
100\% of the sum rule (51).
The dotted curve is the nuclear contribution only and
the solid curve includes both contributions and their interference.
We see that the scattering data are very reasonably described by this
approach. The excitation of isovector electric dipole giant resonance in the
same system is dominated by the Coulomb interaction and is also well
described by  eqs. (47,48), as shown in ref. \cite{BN93}.

In general, both the deformed potential model and the folding model
give very good agreement with the experimental data for the inelastic
scattering of stable nuclei. This has been studied extensively, e.g.,
in ref. \cite{Sa87}. Also, in ref. \cite{BS93} it was shown that a good
agreement with the experimental data for the excitation of giant
monopole and giant quadrupole resonances in lead by 172 MeV $\alpha$'s
is obtained with the use of the eqs. (42-44),
based on the eikonal approximation.
The details of the oscillatory pattern of the angular distributions are
easily understood in terms of the Bessel function
$J_\mu (q_t b)$ in eq. (42).

An interesting application of this formalism is to the excitation of
a Roper resonance ($E_x=500$ MeV) in the $\alpha+p$ reaction with
$E_\alpha=4.2$ GeV.
This is shown in fig. 10, together with
the data points taken  from ref. \cite{Mo92}.
The data show a very steep angular dependence, characteristic of a
monopole transition. As shown by those authors,
the angular distribution can be well described by assuming that the
Roper resonance is a monopole excitation, exhausting a large fraction
of the monopole sum rule, eq. (30).
The solid curve in the figure is calculated by using a folding potential
derived according to eq (17) and the transition amplitude calculated
within the deformed potential model, eq. (39), and using $\delta_0$
calculated as in eq. (30), with $<r^2_p>=3R^2/5=0.69$ $fm^2$ (and $A=1$).
The agreement with the data is remarkably good and the value of
$<r^2_p>$ is in reasonable agreement with the predictions
for the electromagnetic radius of the nucleon
in models which include sea-quark polarization effects \cite{Be73}.

To our knowledge,
up to the present date
very few experimental data exist on the inelastic scattering
of exotic nuclei. Using the formalism in eqs. (40-44) predictions for
the inelastic excitation of soft monopole and quadrupole modes in
the $^{11}Li$ incident on carbon targets has been done \cite{BS93}.
It was shown that the particular oscillatory pattern of the excitation
cross sections could be an important tool to identify different
excited states in these nuclei. This is a well known method in
reactions with stable nuclei, which allows one to distinguish,
e.g., monopole from quadrupole excitations. The extension of the
halo in loosely-bound nuclei would also affect appreciably the
drop in magnitude of these cross sections with increasing scattering
angle \cite{BS93}.

In figure 11 we plot the excitation cross section for the $2^+$ state
($E_x=3.57$ MeV) in $^8He$ by 72 MeV protons. The data are from ref.
\cite{Ko93}. We used the folding model for the calculation of
the optical potential.
The inelastic cross section was calculated with the deformed
potential model and a deformation parameter $\beta_2=0.32$.
As seen, the agreement with the data is quite good.

Finally we discuss the Coulomb excitation of exotic
nuclei. As an example we consider the Coulomb excitation
of $^{11}Be$ projectiles with 45 MeV/nucleon incident on lead targets.
We consider the electric dipole excitation of the
$1/2^+$ ground state to the $1/2^-$  state at
0.32 MeV.
This experiment has been recently done at GANIL \cite{Ha94}
as an initiative of studying the
properties of low-lying states in exotic nuclei.
We use eq. (47) with
$B(E1)=0.116$ $e^2$ $fm^2$ which is deduced from
the lifetime measurement by Millener {\it et al.} \cite{Mi83}.
We obtain the solid curve shown in
fig. 12.
The Coulomb cross section is peaked at very
forward angles, as expected.
The angular integrated cross section is equal to 210 mb.
We omit the nuclear contribution which is very small and only
important for large
scattering angles.
The cross section presents a wiggling
at large scattering angles, characteristic of diffraction patterns
in inelastic scattering. It is instructive to compare this result with
a semiclassical calculations, based on Rutherford orbits for the
trajectory of the nuclei and time-dependent perturbation theory
\cite{BB88}. In this case, the excitation cross section is given
by
\begin{equation}
{d \sigma \over d\Omega} (E_x) = {16\over 9\hbar c} \ \pi^3 \
B(E1, E_x) \ {dn_{E1}\over d\Omega} \ ,
\end{equation}
where the equivalent photon numbers $dn_{E1}/d\Omega$ are given
analytically by
\begin{equation}
{d n_{E1} \over d \Omega_a}={Z_A^2 \alpha \over 4 \pi^2}
 \; \bigl({c\over v} \bigr)^2
 \; {\epsilon^4 \; \zeta^2 }\;
\; e^{-\pi \zeta } \; \biggl\{ {1 \over \gamma^2} \;
{\epsilon^2 -1 \over \epsilon^2} \;
\bigl[ K_{i\zeta} ({\epsilon\zeta}) \bigr]^2
+ \bigl[ K'_{i\zeta} ({\epsilon\zeta})\bigr]^2 \biggr\}
\end{equation}
where $\epsilon=1/\sin (\theta/2)$, $\alpha=1/137$,
$\zeta=E_xa_0/\gamma \hbar v$, with $a_0=Z_aZ_A e^2/2 E_{Lab}$.

Using the above expression for
$d n_{E1} / d \Omega_a$
we obtain the
dashed curve in figure 12 for the same reaction
One observes that the  quantum result
does not deviate from the semiclassical
result appreciably.
The good  agreement especially at lower angles
is quite satisfactory.
This shows that strong absorption
is not relevant for the scattering at low angles.
The peak at small angles is a consequence of the adiabaticity
condition. For $\zeta \gg 1$ ($\theta_a\ll 0.2^\circ$) the Coulomb field
is too weak to provide the necessary excitation energy. For
$\zeta\ll 1$ ($\theta_a\gg 0.2^\circ$) the Coulomb field is
too strong and privileges the
excitation of the projectile
to higher energy states.
Therefore, the cross section
at a fixed relative energy of the fragments in the final channel
has a peak at the optimal scattering angle corresponding to
that energy.  For the  case above this angle is about $0.2^\circ$.

\bigskip\bigskip

\section{Conclusions}

We have studied the applicability of simple concepts from scattering theory
for intermediate energy collisions as a tool to study the
ground state densities and the transition probabilities in reactions with
radioactive nuclei.
The eikonal
approximation together with
the ``t$\rho\rho$'' approximation yields the simple and transparent
formulas. This model gives very reasonable
results for elastic scattering cross sections
at forward angles which can be used to test
the ground state densities of  radioactive nuclei.
The extended nuclear matter in
exotic nuclei is manifest in the magnitude of the elastic cross sections
as well as in the position of the first minimum.

The transition densities from the ground state to continuum states
have the characteristic of an extended tail at large distances from the
nuclear center. As discussed in ref. \cite{BS93} the folding model for
the inelastic scattering cross sections are more appropriate for
the study of reactions with exotic nuclei. In this article we have
presented a few more cases, complimentary to those studied in ref.
\cite{BS93}.  Comparisons with the experimental data
in figs. (9-12) shows that the
formalism presented here is well suited to the study of inelastic
excitation of radioactive beams. The cross section for some cases
predicted here are large and
are well under the experimental possibilities. Of special interest
would be a possible identification of soft multipole modes in such
experiments. This could be accomplished by looking at the
angular pattern of inelastic scattering since different multipolarities
yield quite different patterns.

\newpage
\noindent
{\bf Figure Captions}  \\

Fig. 1 \,\,\,
Elastic scattering cross section of $^{17}O+^{208}Pb$ at 84 MeV/nucl.
The data points are from ref. \cite{Ba88}.
The solid curve is a
calculation with the ``t$\rho\rho$'' approximation.

\bigskip

Fig. 2 \,\,\,
Elastic scattering cross section of $\alpha+\alpha$ at $E_{lab}=2.57$ GeV
as a function of the invariant momentum transfer $t=-(p_1+p_2)^2$.
The data points are from ref. \cite{Be80}.
The solid curve is a
calculation with the ``t$\rho\rho$'' approximation.

\bigskip

Fig. 3 \,\,\,
Elastic scattering cross section of
$^{12}C+^{12}C$ at $E_{lab}=85$ MeV/nucl.
The solid curve uses the eikonal approximation and the
WS optical potential parameters given by eq. (26). The dashed curve
uses the ``t$\rho\rho$'' approximation. The data are taken from ref.
\cite{Bu82}.

\bigskip

Fig. 4 \,\,\,
(a) $p+^9Li$ elastic scattering at $E_p=60$ MeV. Data are from ref.
\cite{Mo92}.
(b) $p+^{11}Li$ elastic scattering at $E_p=62$ MeV. Data are from ref.
\cite{Mo92}.

The solid curves are obtained by using the optical potential as in eqs. (1-3)
with the same set of parameters used in ref. \cite{Mo92}. The dashed curves
use ``t$\rho\rho$'' potentials constructed as in eqs. (17-19).

\bigskip

Fig. 5 \,\,\,
Elastic scattering cross section for $p+^8He$ at 72 MeV/nucl. The solid
curve is a calculation using eqs. (17-19), while the dashed curve
is for the system $p+^6He$ at the same bombarding energy.
The data are from ref.
\cite{Ko93}.

\bigskip

Fig. 6 \,\,\,
Quasi-elastic scattering of $^{11}Li+^{12}C$ at $E_{lab}=637$ MeV.
Data are from ref. \cite{Ko92}.
The inelastic contribution to the
cross section was added
to the elastic cross section (solid figure).
Pure elastic cross section is given by the dashed curve.
The deformation parameters $\beta_2=0.59$
and $\beta_3=0.40$ for the $2^+$ and $3^-$ states in $^{12}C$ were
used.

\bigskip

Fig. 7 \,\,\,
Quasi-elastic scattering of
$^{12}Be+^{12}C$ at $E_{lab}=796$
MeV. Data are from ref. \cite{Za94}.
The inelastic contribution to the
cross section was added
to the elastic cross section (solid figure).
Pure elastic cross section is given by the dashed curve.
The deformation parameters $\beta_2=0.59$
and $\beta_3=0.40$ for the $2^+$ and $3^-$ states in $^{12}C$ were
used.

\bigskip

Fig. 8 \,\,\,
Quasi-elastic scattering of ${14}Be+^{12}C$  at $E_{lab}=679$
MeV. Data are from ref. \cite{Za94}.
The inelastic contribution to the
cross section was added
to the elastic cross section (solid figure).
Pure elastic cross section is given by the dashed curve.
The deformation parameters $\beta_2=0.59$
and $\beta_3=0.40$ for the $2^+$ and $3^-$ states in $^{12}C$ were
used.

\bigskip

Fig. 9 \,\,\,
Excitation cross section of the giant quadrupole resonance in $^{208}Pb$
in the reaction $^{17}O+^{208}Pb$ at 84 MeV/nucleon. Data are from
ref. \cite{Ba88}. The optical potential was calculated by the
``t$\rho\rho$'' approximation.
The deformed potential model is used to calculate the nuclear
excitation cross section  which is drawn by a dotted curve.
The dashed curve is the Coulomb excitation cross
section. The total cross section
due to Coulomb and to nuclear excitations is shown by the solid curve.

\bigskip

Fig. 10 \,\,\,
Angular dependence of the inelastic cross section for the excitation
of the Roper resonance in proton with the reaction $\alpha+p$ at
$E_\alpha=4.2$ GeV. Data points are from ref. \cite{Mor92}.
The solid curve is calculated by using the deformed potential model.

\bigskip

Fig. 11 \,\,\,
Excitation cross section of the $2^+$-state ($E_x=3.57$ MeV) in
$^8He$ by protons with 72 MeV. Data points are from ref. \cite{Ko93}.
The solid curve is obtained by using the ``t$\rho\rho$'' potential
for $p+^8He$ and the deformed potential model for the excitation,
with $\beta_2=0.32$.

\bigskip

Fig. 12 \,\,\,
Coulomb excitation cross section of the $1/2^-$ state at 320 keV in
$^{11}Be$ incident at 45 MeV/nucl. on a lead target.
The solid curve uses the eqs. (47-49), with the eikonal phase calculated
with the ``t$\rho\rho$ approximation. The dashed curve is a semiclassical
calculation using eqs. (52-53).

\newpage

\begin{table}[h]
\noindent
Table 1. Single-neutron and two-neutron separation
energies of neutron-rich nuclei.  Data are taken from
ref. \cite{Wap93}.\\
\vspace{-1.0 mm}
\begin{center}
\begin{tabular}{ll|c|r} \hline
  A& $J^{\pi }$ & $S_{n}$(MeV) & $S_{2n}$(MeV) \\\hline
 $^{6}$He& ${0}^{+}$& 1.86&0.97 \\
 $^{8}$He& ${0}^{+}$& 2.58$\pm$ 0.01&2.14$\pm$ 0.05 \\\hline
 $^{9}$Li& $\frac{3}{2}^{-}$& 4.06&6.10 \\
  $^{11}$Li&  $\frac{3}{2}^{-}$  & 0.73$\pm$ 0.05 & 0.31$\pm$ 0.05 \\\hline
 $^{11}$Be & $\frac{1}{2}^{+}$ & 0.51 &   7.32   \\
 $^{12}$Be & $0^{+}$          & 3.17   &  3.67   \\
  $^{14}$Be & $0^{+}$          & 3.35$\pm$ 0.11   &  1.34$\pm$ 0.11 \\ \hline
 \end{tabular}
\end{center}
\end{table}

\bigskip\bigskip

\begin{table}[h]
\noindent
Table 2. Calculated mass radii and observed interaction radii of halo nuclei.
Radii of some halo orbits are also tabulated in the table. The core radii
are obtained by the H-F calculations, while the values for
the halo configurations are calculated by using the renormalized potential (3).
 Data are taken from refs. \cite{Tan85,Tan88}.
\vspace{2.0 mm}
\begin{center}
\begin{tabular}{ll|c|r} \hline
  A & j$_{last}$  &$\sqrt{\langle r^{2} \rangle_{cal}}$ (fm)&
     $\sqrt{\langle r^{2} \rangle_{exp}}$  (fm)\\\hline
 $^{6}$He &   &2.66  &2.48$\pm$ 0.03\\
 $^{8}$He &     &2.59  &2.49$\pm$0.03   \\\hline
 $^{9}$Li &            &2.45  &2.41$\pm$ 0.02\\
 $^{11}$Li& 1p$_{1/2}$&5.36  &               \\
             &         &3.08   &3.20$\pm$ 0.03 \\\hline
 $^{11}$Be & 2s$_{1/2}$& 6.29&        \\
      &        & 3.01  &  2.86$\pm$ 0.04   \\\hline
 $^{12}$Be & 2s$_{1/2}$& 3.46&        \\
      &        & 2.68 &  2.82$\pm$ 0.04   \\\hline
 $^{14}$Be & 2s$_{1/2}$& 5.28&        \\
      &        & 3.16  &  3.33$\pm$ 0.17   \\\hline
\end{tabular}
\end{center}
\end{table}

\bigskip
\bigskip

\newpage
Table 3.  Parameters of the ground states densities of stable nuclei
which are given by either gaussian (G) or modified fermi (MF)
distributions.  Data are taken from \cite{VJV87}.

\begin{center}
\begin{tabular}{|l|l||l|l|l|l|l|r|} \hline\hline
Nucleus&Model&R&a&c \\ \hline
$\alpha$&Gaussian&1.37&--&-- \\ \hline
$^{11}C$&MF&2.34&0.5224&-0.149 \\ \hline
$^{12}C$&MF&2.34&0.5224&-0.149 \\ \hline
$^{16}O$&MF&2.61&0.513&-0.051 \\ \hline
$^{17}O$&MF&2.61&0.513&-0.051 \\ \hline
$^{208}Pb$&MF&6.62&0.549&0 \\ \hline
\end{tabular}
\end{center}

\bigskip
\bigskip

Table 4.
Parameters \cite{Ra79} for the nucleon-nucleon amplitude, $f_{NN}(q)=
(k_{NN}/4\pi) \ \sigma_{NN} \ (i+\alpha_{NN})\ e^{-\xi_{NN}q^2} $.

\begin{center}
\begin{tabular}{|l|l||l|l|l|l|l|r|} \hline\hline
E [MeV/nucl]&$\sigma_{NN}$ [fm$^2$]&$\alpha_{NN}$&$\xi_{NN}$ [fm$^2$] \\ \hline
30&19.6&0.87& \\ \hline
38&14.6&0.89& \\ \hline
40&13.5&0.9& \\ \hline
49&10.4&0.94& \\ \hline
85&6.1&1& \\ \hline
94&5.5&1.07&0.51 \\ \hline
120&4.5&0.7&0.58 \\ \hline
200&3.2&0.6&0.62 \\ \hline
342.5&2.84&0.26&0.31 \\ \hline
425&3.2&0.36&0.24 \\ \hline
550&3.62&0.04&0.062 \\ \hline
650&4.0&-0.095&0.08 \\ \hline
800&4.26&-0.075&0.105 \\ \hline
1000&4.32&-0.275&0.105 \\ \hline
2200&4.33&-0.33&0.13 \\ \hline
\end{tabular}
\end{center}
\bigskip
\bigskip

\end{document}